\documentclass[jkps,twocolumn,fleqn,showpacs,showkeys,floatfix]{revtex4}
\usepackage{graphicx}
\usepackage{amssymb}
\usepackage{amsmath}
\usepackage{bm}
\begin{document}
\title[]{
Estimate of the Phase Transition Line in the Infinite-dimensional Hubbard Model
}
\author{Aaram J. \surname{Kim}}
\author{M.Y. \surname{Choi}}
\affiliation{Department of Physics and Astronomy and Center for Theoretical Physics, Seoul National University, Seoul 151-747}
\author{Gun Sang \surname{Jeon}}
\email{gsgeon@ewha.ac.kr}
\affiliation{Department of Physics, Ewha Womans University, Seoul 120-750}

\date[]{}

\begin{abstract}
We consider a Mott transition of the Hubbard model in infinite
dimensions.
The dynamical mean-field theory is employed in combination with a
continuous-time quantum Monte Carlo (CTQMC) method for an accurate description
at low temperatures.  
From the double occupancy and the energy density, which are  directly measured from the
CTQMC method, we construct the phase diagram.
We pay particular attention to the construction of the first-order phase
transition line (PTL) in the coexistence region of metallic and insulating phases.
The resulting PTL is found to exhibit reasonable
agreement with earlier finite-temperature results.
We also show by a systematic inclusion of low-temperature data that the PTL, which is achieved independently of the previous zero-temperature results, approaches monotonically the transition point from earlier 
zero-temperature studies. 
\end{abstract}



\maketitle

\section{Introduction}
Over the past few decades, the dynamical mean-field
theory (DMFT)~\cite{Metzner1989,Georges1996} and its extensions~\cite{Maier2005}
have proven to be successful in describing the dynamic properties of
strongly correlated systems. 
In DMFT, a lattice model is mapped onto a corresponding quantum impurity model
such as the Anderson impurity model.
The impurity self-energy, which is obtained by solving the quantum impurity
model, is identified with that of the lattice model, 
imposing a self-consistency relation.
Here, we assume that the self-energy is local; this is the case in infinite
spatial dimensions and provides a good approximation in high dimensions.

The efficiency of the DMFT depends strongly on the method of solving the quantum
impurity problem, which is called {\em an impurity solver}.
Some perturbative methods such as iterative perturbation
theory~\cite{Georges1992a} and the noncrossing approximation~\cite{Pruschke1993}
were proposed and turned out to yield qualitatively correct results.  
For a quantitatively accurate description, however, 
nonperturbative methods must be developed.
One of the reliable nonperturbative tools is the Hirsch-Fye quantum Monte Carlo
method~\cite{Hirsch1986}, which is useful at finite temperatures.
In that method, however, the discretization of the imaginary-time axis raises
difficulties in capturing the sharp variation of the imaginary-time
Green function, particularly at low temperatures.
Such a problem can be resolved by
using the recently-developed continuous-time quantum Monte Carlo (CTQMC) method, 
which performs a stochastic sampling of an expansion order in the imaginary-time 
axis without any discretization~\cite{Rubtsov2005,Gull2008a,Werner2006,Gull2011}.

We here intend to examine the finite-temperature Mott transition in 
the Hubbard model, which is well known to exhibit the characteristic features of
the Mott transition~\cite{Mott1949,Gebhard}.
Previous extensive studies help us to understand the qualitative nature of the
Mott transition in the Hubbard model:
There exists a critical temperature above which only a
smooth crossover occurs between a metal and a Mott insulator~\cite{Rozenberg1999}. 
Below the critical temperature, in contrast, a coexistence region separates the metallic from the Mott insulating phase~\cite{Rozenberg1999,Bulla2001},
which implies that the system undergoes
a first-order phase transition somewhere inside the coexistence
region~\cite{Rozenberg1994a,Georges1996}.

Our work focuses on the numerical determination of a phase transition line (PTL) of the
first-order Mott transition.
We can determine a thermodynamically-stable phase from the criterion of the
lowest free energy. Also, at finite temperatures, the entropy makes
important contributions.
Although the QMC method is very useful at finite temperatures,
obtaining the entropy directly from the QMC method is very difficult.
Accordingly,
numerical computation of the PTL has been a challenging
problem.
Only a few earlier studies here made efforts to obtain the PTL by using exact
diagonalization~\cite{Tong2001} or Hirsch-Fye QMC method~\cite{Blumer2002}, and a recent
CTQMC study succeeded in locating a single transition point at
a very low temperature~\cite{Werner2006}.

In this paper, we present a reliable numerical scheme for estimating
the PTL of the Mott transition at low temperatures.
We basically use the same differential equation as in
Ref.~\onlinecite{Blumer2002}.
However, in contrast to that previous study, where some fitting functions were introduced~\cite{Blumer2002},
we measure the quantities for the equation directly by using the CTQMC method without any additional manipulation.
We show that our method results in the PTL which is fully consistent with 
earlier finite-temperature studies. 
We also demonstrate that our finite-temperature result approaches gradually the
zero-temperature result when we systematically include low-temperature
calculations.
Finally, the resulting PTL is found not to be changed significantly by the
next-order correction to some approximations that are made while solving the
differential equation.

This paper is organized as follows: Section~II is devoted to
the description of the Hubbard model and the DMFT combined with the CTQMC method.
We present the results in Sec.~III: 
We display some physical quantities, such as the double occupancy and the energy density, 
from the CTQMC method, and we present a phase diagram constructed from the data.
Particularly, we describe how we can construct the PTL from the QMC data
and give some discussion on the resulting PTL.
The results are summarized in Sec.~IV.

\section{Model and Method}
\subsection{One-band Hubbard Model}
We begin with the one-band Hubbard model.
Its Hamiltonian is given by
\begin{equation}
	\mathcal{H} = -t\sum_{\langle ij\rangle\sigma}(\hat{c}_{i\sigma}^\dagger \hat{c}_{j\sigma} + \hat{c}_{j\sigma}^\dagger \hat{c}_{i\sigma} ) + U\sum_i \hat{n}_{i\uparrow}\hat{n}_{i\downarrow} - \mu\sum_{i\sigma} \hat{n}_{i\sigma}~.
	\label{Hamiltonian}
\end{equation}
Here, $\hat{c}^{}_{i\sigma}$ ($\hat{c}_{i\sigma}^\dagger$) is an annihilation (creation) operator for an electron with spin $\sigma$ at site $i$, and $\hat{n}_{i\sigma} \equiv \hat{c}_{i\sigma}^\dagger \hat{c}^{}_{i\sigma}$.
The parameters $t$ and $U$ denote the nearest-neighbor hopping amplitude and 
the on-site Coulomb repulsion, respectively.
The chemical potential $\mu$ is set to $U/2$ so that the system is half-filled. 
We consider a Bethe lattice in infinite dimensions, which results in 
a semicircular density of state 
\begin{equation}
\label{eq:DOS}
  \rho(\varepsilon) = \frac2{\pi D}\sqrt{1 - (\varepsilon/D)^2}
\end{equation}
with a half bandwidth $D{=}2t$.
We restrict our study to paramagnetic solutions, and throughout the paper, we will represent all the energies in units of $D$.

\subsection{Dynamical Mean-field Theory}
The single-site DMFT, which we will employ in our work, has been known to be a very efficient tool for investigating strongly correlated systems~\cite{Georges1996}.
It incorporates all the local quantum fluctuations that play an important role in strongly correlated systems.  
The central idea of the DMFT is to map a lattice model onto a quantum impurity model, which should satisfy a self-consistency relation imposed from the original lattice model.
The effective action of the quantum impurity model is written as
\begin{eqnarray} \label{eq:Seff}
	S_\mathrm{eff} &=& S_o~\nonumber\\
		&&+ \sum_\sigma\int^{\beta}_{0}d\tau\int^{\beta}_{0}d\tau'~c_\sigma^{\dagger}(\tau)\Delta_\sigma(\tau - \tau')c^{}_{\sigma}(\tau')~,\nonumber\\
	\label{eq:effaction}
\end{eqnarray}
where $S_o$ is an action at the impurity site due to the local interaction, as well as the chemical potential, and $\beta{\equiv} 1/T$ is an inverse temperature.
The hybridization function $\Delta_\sigma(\tau)$ in Eq.~(\ref{eq:Seff}) plays the role of a generalized Weiss field. 
We should note that it is a function of the imaginary time $\tau$.

In infinite dimensions the self-energy is local, that is, independent of the momentum $\bm{k}$, and
the local Green function with spin $\sigma$ is given by
\begin{equation}
	G_\sigma(i\omega_n) = \int_{-\infty}^{\infty} d\varepsilon \frac{\rho(\varepsilon)}{i\omega_n - \varepsilon +\mu - \Sigma_\sigma(i\omega_n)}.
\end{equation}
The corresponding hybridization function $\Delta_\sigma$ can also be calculated as 
\begin{equation}
	\Delta_\sigma(i\omega_n) =  i\omega_n + \mu - \Sigma_\sigma(i\omega_n)- G^{-1}_\sigma(i\omega_n) .
\end{equation}
In the DMFT, the local self-energy is assumed to be an impurity self-energy, yielding 
\begin{eqnarray}
\lefteqn{
\left[\int_{-\infty}^{\infty} d\varepsilon \frac{\rho(\varepsilon)}{i\omega_n - \varepsilon +\mu - \Sigma_\sigma(i\omega_n)}\right]^{-1}
} \phantom{aaaaaaaaaa}
\nonumber
\\
&	= & i\omega_n + \mu - \Sigma_\sigma(i\omega_n)-\Delta_\sigma(i\omega_n) .
\label{eq:gselfconsistency}
\end{eqnarray}
This is the self-consistency relation that should be satisfied by the 
self-energy of the quantum impurity with the hybridization function $\Delta_\sigma$.
By using a semicircular density of states for $\rho(\epsilon)$ in Eq.~(\ref{eq:DOS}), 
	we can further simplify
the above self-consistency relation to the form
\begin{equation}
\Delta_\sigma (\tau) = \frac{D^2}{4} G_\sigma(\tau)
\end{equation}
in the infinite-dimensional Bethe lattice. 

\subsection{Continuous-time Quantum Monte Carlo Method}
In order to solve a quantum lattice problem within the DMFT, we need an impurity solver that gives an impurity self-energy for a given hybridization function.
In this work, we employ the CTQMC method as the impurity solver.
In particular, we use the hybridization expansion algorithm,
which allows an accurate calculation even for strong interactions, as well as very low temperatures. 
Within the algorithm, we can also
measure the kinetic energy density directly from the Monte Carlo sampling. 

The basic idea of the QMC method is the stochastic sampling of perturbation diagrams of a partition function.  
In the system with the Hamiltonian $\mathcal{H} {=} \mathcal{H}_a {+} \mathcal{H}_b$, we can express the partition function as
\begin{equation}
	Z = \mathrm{Tr} \left[e^{-\beta \mathcal{H}_a}T_\tau\exp\left(-\int^\beta_0 d\tau \tilde{\mathcal{H}}_b(\tau)\right)\right]~,
	\label{eq:Z}
\end{equation}
where $\tilde{\mathcal{O}}(\tau) {\equiv} e^{\tau \mathcal{H}_a}\mathcal{O} e^{-\tau\mathcal{H}_a}$ for an operator $\mathcal{O}$ and $T_\tau$ represents a time-ordering operator.
In the CTQMC method,
a random walker roams the configuration space composed of a perturbation order, perturbation positions on the imaginary-time axis, and other parameters depending on the specific algorithm.

The Hamiltonian of the Anderson impurity model can be expressed as 
\begin{equation}
\mathcal{H} = \mathcal{H}_\mathrm{bath} + \mathcal{H}_\mathrm{hyb} + \mathcal{H}_\mathrm{loc}, 
\end{equation}
where
\begin{eqnarray}
	\mathcal{H}_\mathrm{bath} &\equiv& \sum_{\bm{p}\sigma} \varepsilon_{\bm{p}} \hat{a}^{\dagger}_{\bm{p}\sigma} \hat{a}^{}_{\bm{p}\sigma}~,
\\
	\mathcal{H}_\mathrm{hyb} &\equiv& \sum_{\bm{p}\sigma}~(V_{\bm{p}\sigma} \hat{a}^\dagger_{\bm{p}\sigma}\hat{c}^{}_\sigma + V^{*}_{\bm{p}\sigma }\hat{c}^\dagger_\sigma \hat{a}^{}_{\bm{p}\sigma})~,
	\\ 
	\mathcal{H}_\mathrm{loc} &\equiv& U\hat{n}_{\uparrow} \hat{n}_{\downarrow} - \mu \sum_{\sigma}\hat{n}_\sigma.
\end{eqnarray}
In the hybridization expansion algorithm, we take 
$\mathcal{H}_\mathrm{bath} {+} \mathcal{H}_\mathrm{loc}$ 
as an unperturbed Hamiltonian $\mathcal{H}_a$ 
and expand the full Hamiltonian 
to the order of $\mathcal{H}_b {=} \mathcal{H}_\mathrm{hyb}$.
By recalling that $\mathcal{H}_\mathrm{hyb} {=} \sum_\sigma(h^{}_{\sigma} {+} h^\dagger_{\sigma})$
with
$h_\sigma {\equiv} \sum_{\bm{p}}V_{\bm{p}\sigma} \hat{a}^\dagger_{\bm{p}\sigma}\hat{c}^{}_\sigma$,
we can rewrite Eq.~(\ref{eq:Z}) in the form 
\begin{eqnarray}
	Z &=& \mathrm{Tr}\Bigg[e^{-\beta \mathcal{H}_{a}}\prod_{\sigma}\sum^\infty_{k_\sigma=0}\nonumber\\
	&\times&\int^\beta_0d\tau_{1_\sigma}\int^\beta_0d\tau_{1_\sigma}'\cdots\int^\beta_{\tau_{k_\sigma-1}}d\tau_{k_\sigma}\int^\beta_{\tau_{k_\sigma-1}'}d\tau_{k_\sigma}'\nonumber\\
	&\times&\tilde{h}_{\sigma}(\tau_{k_\sigma})\tilde{h}^\dagger_{\sigma}(\tau'_{k_\sigma})\cdots \tilde{h}_{\sigma}(\tau_{1_\sigma})\tilde{h}^\dagger_{\sigma}(\tau'_{1_\sigma})\Bigg]~.
\end{eqnarray}
Here, we have used the fact that only the terms in which $\tilde{h}_{\sigma}$ and $\tilde{h}^\dagger_{\sigma}$ appear alternately the same number of times produce nonzero traces 
due to the fermionic nature of electrons.

We use a bath partition function $Z_\mathrm{bath}$ defined by
\begin{equation}
	Z_\mathrm{bath} \equiv \mathrm{Tr}_a[e^{-\beta \mathcal{H}_\mathrm{bath}}] = \prod_{\bm{p}\sigma}(1+e^{-\beta \varepsilon_{\bm{p}}})
\end{equation}
and apply the Wick theorem to the bath fermionic operators
to obtain the expanded form of the partition function
\begin{widetext}
	\begin{equation}
		Z = Z_\mathrm{bath}\mathrm{Tr}_{c}\left[e^{-\beta \mathcal{H}_\mathrm{loc}}
			\prod_\sigma\sum_{k_\sigma=0}^{\infty}
			\int^{\infty}_0d\tau_{1_\sigma}\int^{\infty}_0d\tau_{1_\sigma}'\cdots 
			\int^\infty_{\tau_{k_\sigma-1}}d\tau_{k_\sigma}
			\int^\infty_{\tau_{k_\sigma-1}'}d\tau'_{k_\sigma}~			
			\tilde{c}^{}_\sigma(\tau_{k_\sigma})\tilde{c}_\sigma^\dagger(\tau'_{k_\sigma})
			\cdots 
			\tilde{c}^{}_\sigma(\tau_{1_\sigma})\tilde{c}_\sigma^\dagger(\tau'_{1_\sigma})
			\det\bm{\Delta}_\sigma\right].
	\end{equation}
\end{widetext}
Here, $\bm{\Delta}_\sigma$ is a $k_\sigma \times k_\sigma$ matrix with elements
$$
(\bm{\Delta}_\sigma)_{ij} \equiv \Delta_\sigma(\tau_{i_\sigma}' -
\tau_{j_\sigma}),
$$
with the antiperiodic hybridization function being given by
\begin{equation}
	\Delta_{\sigma}(\tau) = 
	[\theta(-\tau)-\theta(\tau)] 
	\sum_{\bm{p}}\frac{|V_{\bm{p}\sigma}|^2}{e^{\beta\varepsilon_{\bm{p}}} + 1}
	e^{-\{\tau-\beta\theta(\tau)\}\varepsilon_{\bm{p}}}
\end{equation}
and with $\theta(\tau)$ being a step function.

In the Anderson impurity model, the occupation number operator at the impurity
commutes with $\mathcal{H}_a$, so we can give a systematic description of the local trace factor 
in terms of {\em segments} and {\em antisegments};
they represent time intervals during which an electron is present and absent at the impurity, respectively.
In the segment description, the local weight factor for
$x {=} \prod_\sigma\{(\tau_{1_\sigma},\tau_{1_\sigma}'),\cdots,(\tau_{k_\sigma},\tau_{k_\sigma}')\}$
is given by

\begin{eqnarray}
	\omega_\mathrm{loc}(x) &\equiv& \mathrm{Tr}_c \Big[e^{-\beta\mathcal{H}_{loc}}\nonumber\\
	       && \times \prod_\sigma\tilde{c}^{}_\sigma(\tau_{k_\sigma})\tilde{c}_\sigma^\dagger(\tau'_{k_\sigma})
\cdots 
\tilde{c}^{}_\sigma(\tau_{1_\sigma})\tilde{c}_\sigma^\dagger(\tau'_{1_\sigma})
\Big]\nonumber\\
&=& se^{\mu\sum_{n\sigma}L_{n\sigma}}e^{-U \sum_{nm}O_{nm}}~,
\end{eqnarray}
where $L_{n\sigma}$ is the length of the $n$th segment of an electron with spin $\sigma$, 
$O_{nm}$ is the overlap between the $n$th spin-up and the $m$th spin-down segments 
on the imaginary-time axis, and $s$ is a sign determined by the sequence of operators.
The main procedures in updating the configuration 
are the insertion and the removal of segments or antisegments.
We have separated the proposal probability so that we can get rid of the factor $d\tau^2$ in the weight factor.
We have also used self-balance and global updates, which can reduce the autocorrelation time significantly.

The above procedures enable us to evaluate
the impurity Green function, $G_\sigma(\tau) {\equiv} -\langle T_\tau \hat{c}^{}_\sigma(\tau)\hat{c}^\dagger_\sigma(0)\rangle$, by using the formula
\begin{equation}
	G_{\sigma}(\tau) =
-\frac{1}{\beta}\left\langle\sum^{k_\sigma}_{ij}(\bm{M}_\sigma)_{ji}\tilde{\delta}(\tau,\tau_i
- \tau'_j)\right\rangle,
\end{equation}
where the angular brackets denote the Monte Carlo average, $\bm{M}_\sigma \equiv \bm{\Delta}_\sigma^{-1}$, and

\begin{equation}
	\tilde{\delta}(\tau,\tau') 
	\equiv [\theta(\tau')-\theta(-\tau')] 
	\delta\big(\tau-\tau'- \beta\theta(-\tau')\big).
\end{equation}
At the final stage, we obtain a paramagnetic Green function $G$ 
by symmetrizing the Green function,  
$G {=} (G_\uparrow + G_\downarrow)/2$. 

One advantage of the hybridization expansion is that 
we can measure directly some local quantities during Monte Carlo samplings.
The average occupancy can be evaluated by using the ratio of the Monte Carlo average 
of the total segment length to the length of the imaginary-time axis:
\begin{equation}
	\langle \hat{n}_\sigma \rangle =\frac{1}{\beta}
	\left \langle \sum_n L_{n\sigma} \right\rangle.
\end{equation}
Similarly, the Monte Carlo average of the total overlap length 
gives the double occupancy through the formula
\begin{equation} \label{eq:dO}
	d_O \equiv \langle \hat{n}_\uparrow \hat{n}_\downarrow \rangle
	= \frac{1}{\beta} \left\langle \sum_{nm}O_{nm} \right\rangle.
\end{equation}
In the hybridization expansion, we can also use
the average perturbation order $\langle k_\sigma \rangle$ to evaluate 
the average kinetic energy of electrons with spin $\sigma$, $\varepsilon_{K\sigma}$~\cite{Haule2007,Gull2011}.
We can show that the perturbation order estimates the average value of 
the perturbing action, 
which enables us to obtain the kinetic energy 
directly from the Monte Carlo average of the perturbation order:
\begin{eqnarray}
	\langle k_\sigma\rangle &=& 
 	-\int^\beta_0d\tau\int^\beta_0d\tau'
 	\Delta_\sigma(\tau-\tau')\langle c^{\dagger}_\sigma(\tau)c^{}_\sigma(\tau')\rangle\nonumber\\
	&=& -\mathrm{Tr}[\Delta_\sigma G_\sigma]\nonumber\\
	&=& 
	-\beta \varepsilon_{K\sigma}~.
	\label{eq:avek}
\end{eqnarray}

\section{Results}
\subsection{Double Occupancy and Energy Density}
\begin{figure}
	\includegraphics[width=0.4\textwidth]{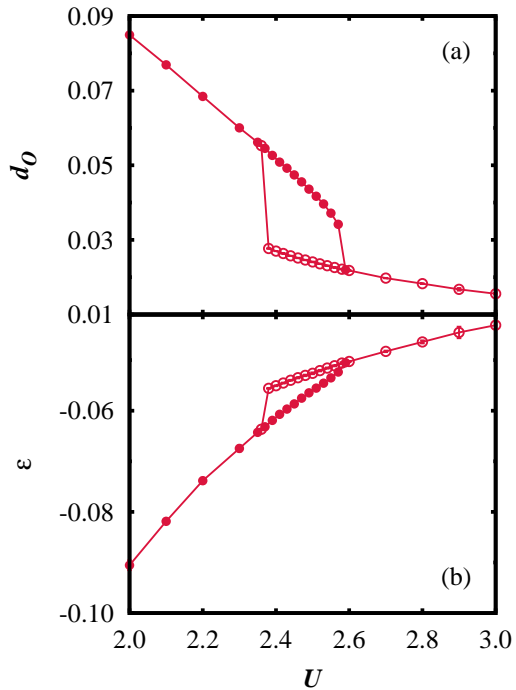}
	\caption{(color online) (a) Double occupancy $d_O$ and (b) energy density $\varepsilon$ as functions of $U$ at the temperature $T {=} 1/128$.  
		The data from the DMFT solutions obtained by increasing and decreasing $U$ are marked by solid and empty circles, respectively. 
	}
	\label{fig:doE}
\end{figure}

We first examine the double occupancy $d_O$ defined in Eq.~(\ref{eq:dO}).
The double occupancy is well known to provide a good measure for the degree of correlation.
As was explained in the previous section, in the CTQMC method, it can be calculated directly from the total overlap length of the segments.
Figure~\ref{fig:doE}(a) presents the calculated double occupancy at the temperature $T{=}1/128$.
As $U$ increases, the system becomes more correlated, and the double occupancy decreases.
At a certain interaction strength $U_{c2}(T)$, the double occupancy shows a discontinuous jump to a lower value, and the system becomes insulating.
On the other hand, when the interaction strength is decreased from that for the insulating solutions, the system exhibits a discontinuous jump in the double occupancy at the interaction strength $U_{c1}(T)$, which is lower than $U_{c2}(T)$.
Accordingly, we have a finite region $U_{c1}{<}U{<}U_{c2}$ where both metallic and insulating phases coexist; 
this demonstrates clearly the first-order nature of 
the metal-insulator transition in the Hubbard model for infinite dimensions.
Such a first-order nature of the transition is also demonstrated in the variation of the energy density. 
We have calculated the energy per lattice site $\varepsilon$ as
\begin{equation}
\varepsilon = \varepsilon_K + \varepsilon_U,
\end{equation}
where we can compute the kinetic energy $\varepsilon_K$ and the interaction energy $\varepsilon_U$
directly from the average quantities in the CTQMC method:
\begin{eqnarray}
	\varepsilon_K &\equiv&  -T\sum_\sigma\langle k_\sigma \rangle ,
\\
\varepsilon_U &\equiv&  U d_O.
\end{eqnarray}
We observe a clear hysteresis between the metallic and the insulating solutions in the energy density as in the double occupancy.  
In the coexistence region, the metallic phase always has a lower energy density than the insulating phase at $T{=}1/128$, which is consistent with earlier DMFT results~\cite{Georges1996,Bulla1999}.

\subsection{Phase Diagram and Critical Point}

\begin{figure}
	\includegraphics[width=0.4\textwidth]{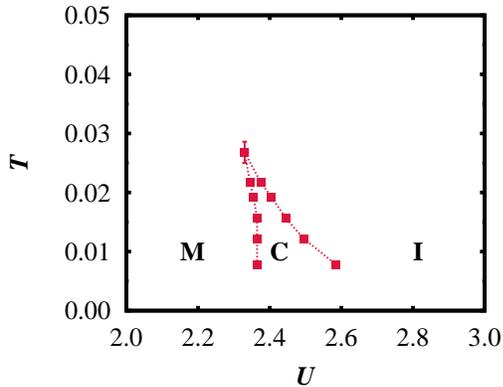}
	\caption{ \label{fig:pd}
		(color online) Phase diagram for the Mott transition in the infinite-dimensional Hubbard model.
		The transition interaction strengths $U_{c1}$ and $U_{c2}$ (see
		the text for definitions) are plotted for various temperatures,
		and the lines are merely guides to the eye. 
		The regions of the metallic and the insulating phases and their
		coexisting regions are denoted by the
		labels M, I, and C, respectively.
	}
\end{figure}

We can determine two transition points, $U_{c1}(T)$ and $U_{c2}(T)$, from the interaction
strengths at which the double occupancy or the energy density shows
discontinuous jumps at temperature $T$. 
In Fig.~\ref{fig:pd} we plot $U_{c1}$ and $U_{c2}$ as functions of the temperature $T$, 
displaying the phase diagram for metal-insulator
transitions on the plane of $T$ and $U$.
Both  $U_{c1}$ and $U_{c2}$ increase monotonically as the temperature is lowered.
With decreasing $T$, the rate of increase of $U_{c1}$ diminishes while it is enhanced for $U_{c2}$. 
The two transition lines merge at the critical temperature $T_c$, which gives an upper bound
on the temperature of the coexistence region denoted by C in Fig.~\ref{fig:pd}.
Above $T_c$, insulating and metallic phases are connected gradually without any
abrupt change.

\begin{figure}
	\includegraphics[width=0.40\textwidth]{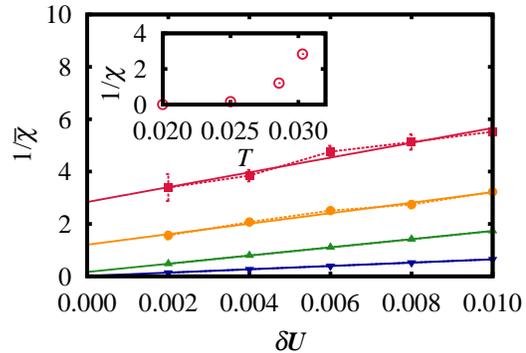}
	\caption{(color online) 
		The inverse susceptibility $1/\bar{\chi}$ estimated by using numerical
		derivatives with a finite interval $\delta U$ for various
		temperatures.  
		The data for $T=1/33, 1/35, 1/40$, and $1/50$ are marked by (red) squares, (orange) circles,
	(green) triangles, and (blue) inverted triangles, respectively (from top
to bottom).  The solid lines represent the best linear fits for each temperature.  
The inset shows the inverse susceptibility $1/\chi$, which is estimated by
extrapolating to  $\delta U {=} 0$.}
	\label{fig:Tc}
\end{figure}

By using the liquid-gas analogy~\cite{Castellani1979},
we define the susceptibility~\cite{Rozenberg1999} as 
\begin{equation}
	\chi \equiv \lim_{\delta U \rightarrow 0} \bar{\chi}(\delta U)~,
\end{equation}
with
\begin{equation}
	\bar{\chi}(\delta U) \equiv \max_U \left(-\frac{ d_O(U+\delta U)- d_O
	(U)}{\delta U}\right).
\end{equation}
Because the susceptibility $\chi$ diverges at the critical temperature $T_c$,
we can use $\chi$ to estimate the critical point precisely.
In Fig.~\ref{fig:Tc}, we plot the inverse of $\bar{\chi}$ as a function of
$\delta U$.
We can then estimate $1/\chi(T)$ by extrapolating the best linear fit to 
the data at the temperature $T$ to the $y$-axis.
As is demonstrated in the inset of Fig.~\ref{fig:Tc},
the critical temperature $T_c$ is estimated to be in the range 
$1/40 < T_c < 1/35$, which is consistent with earlier 
results from DMFT combined with QMC~\cite{Rozenberg1999,Blumer2002} and the exact diagonalization~\cite{Tong2001}.
It is slightly smaller than the estimate from the DMFT combined with
the numerical renormalization group (NRG)~\cite{Bulla2001}.
The corresponding critical interaction strength $U_c(T_c){\equiv} U^*$ 
is $U^* {=} 2.33 \pm 0.01$.

\subsection{Phase Transition Line}

In the thermodynamic limit,
the system in equilibrium resides in the phase with the lowest free energy, and the phase transition between the two coexisting phases occurs 
on the line where the free energies of the two phases are the same.
In this section, we will construct the PTL inside the
coexistence region by means of the DMFT combined with the CTQMC method, which will be
sketched below.

The free energy density $f$ of the system is defined by
\begin{equation}
	f \equiv \varepsilon - Ts ,
\end{equation}
where $s$ is the entropy density, and the change in the free energy density is
given by
\begin{equation}
	df = -sdT + d_O dU.
\end{equation}
Although this form is a standard one, it is not useful in our method because computing the entropy by using the QMC method is very difficult.
Recalling the relation
\begin{equation}
	\frac{\partial(\beta f)}{\partial\beta}\Bigg\vert_U 
	= \varepsilon~,
	\label{eq:diff_bf}
\end{equation}
we can consider an alternative form
\begin{equation}
	d(\beta f) = \varepsilon  \ d\beta + \beta  \ d_O \ dU,
\end{equation}
which is convenient to use because we can obtain the energy density directly
from the QMC average.

In the coexistence region,
we take the difference in the free energy densities between the metallic and
the insulating phases, $\Delta f {\equiv} f_\textrm{M} - f_\textrm{I}$, 
and the change of that difference is given by
\begin{equation}
	d(\beta\Delta f) = \Delta \varepsilon \ d\beta + \beta~\Delta d_O \ dU,
\end{equation}
where the subscripts M and I denote the quantities of the metallic and the insulating
phases, respectively.  
On the PTL, the free energies of the two phase are the same, $\Delta f{=}0$; accordingly, the change in $\beta \Delta f$ vanishes when the parameters
change along the PTL, yielding
\begin{eqnarray}
\lefteqn{
	d[\beta\Delta f(\beta,U_c(\beta))]
}\\
&& = \Delta \varepsilon(\beta,U_c(\beta))d\beta + \beta\Delta d_O(\beta,U_c(\beta))dU_c(\beta)
= 0,~\nonumber
\end{eqnarray}
where $U_c(\beta)$ is the interaction strength at which the thermodynamic
transition occurs at the temperature $T{=}1/\beta$.
The resulting differential equation for the PTL is 
\begin{equation}
	\frac{dU_c(\beta)}{d\beta} = -\frac{\Delta \varepsilon(\beta,U_c(\beta))}{\beta\Delta d_O(\beta,U_c(\beta))},
\end{equation}
which can be transformed to a differential equation in temperature $T$:
\begin{equation}
	\frac{dU_c(T)}{dT}= F(T,U_c(T))
	\label{eq:diffeq}
\end{equation}
 with
\begin{equation}
	F(T,U) = \frac{\Delta \varepsilon(T,U)}{T\Delta d_O(T,U)}.
\label{eq:ffunction}
\end{equation}
In principle, the integration of Eq.~(\ref{eq:diffeq}) with an initial condition $U_c(T_c)=U^*$ yields the PTL.

\begin{figure}
	\includegraphics[width=0.40\textwidth]{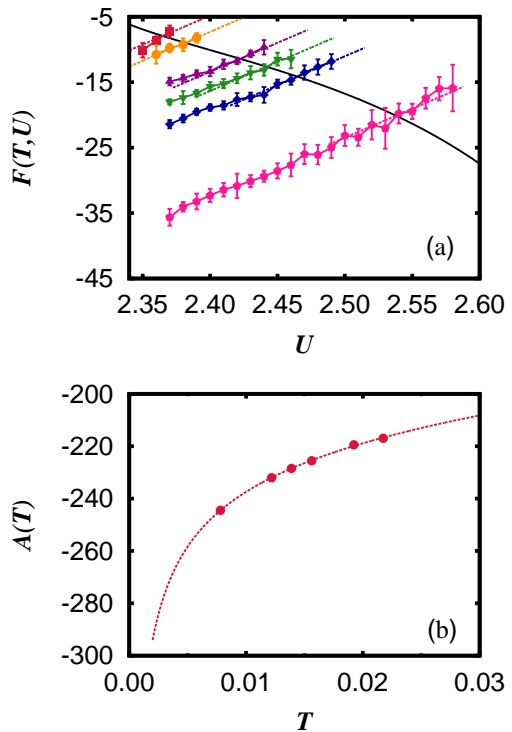}
	\caption{(color online) (a) $F(T,U)$ as a function of the interaction strength $U$ for
		various temperatures 
		and (b) $A(T)$ as a function of the temperature $T$. 
		In panel (a), the data for $T{=}1/46,1/52,1/64,1/72,1/82$, and $1/128$  
		are marked by (red) squares, (orange) circles,
		(violet) triangles, (green) inverted triangles, (blue) diamonds,
		and (pink) pentagons, respectively (from top to bottom). 
		The curve of $(U_c(T),F(T,U_c(T)))$ is also presented by the (black) solid line.
		In panel (b), the dashed line represents the best-fit curve to
		Eq.~(\ref{eq:AT}).
	}
	\label{fig:fA}
\end{figure}

Indeed, Eq.~(\ref{eq:diffeq}) was employed to study the PTL by using the DMFT with the Hirsch-Fye QMC method~[\onlinecite{Blumer2002}]. 
In this study, however,  some model-specific fitting functions were introduced
in evaluating $F(T,U)$ to manage inevitable Trotter errors and the lack of data points,
which may limit the applicability of the method.
In contrast, within the CTQMC method employed in this work, such Trotter errors are absent.
Furthermore, the kinetic energy and the double occupancy can be measured directly from Monte
Carlo sampling.
Thus, we can prepare $F(T,U)$ for Eq.~(\ref{eq:diffeq}) from the raw data obtained with the  CTQMC method by using Eq.~(\ref{eq:ffunction}) without any additional treatment. 

In Fig.~\ref{fig:fA}(a), we plot $F(T,U)$ calculated directly by using the CTQMC method for
various temperatures. 
We can observe that around the PTL, the data for all temperatures agree well with the function
\begin{equation}
	\label{eq:linearU}
	F(T,U) = A(T) + bU
\end{equation}
with $b{=}88.4$.
These functions are denoted by the linear dotted lines in Fig.~\ref{fig:fA}(a).
We have obtained the value of $b$ from the best linear fit at $T{=}1/82$.
The constant $A(T)$ is also obtained from the intercept of the fitting line at temperature $T$ on $U{=}0$ axis.
According to the Fermi-liquid theory, $A(T)$ can be approximated by using the three leading-order terms,
\begin{equation}
	\label{eq:AT}
	A(T) = \frac{\alpha}{\sqrt{T}} + \gamma + \eta \sqrt{T}~,
\end{equation}
at low temperatures,
where we can determine $\alpha, \gamma$ and $\eta$ from the least-square fits.
The integration of Eq.~(\ref{eq:diffeq}) yields an analytic expression for the PTL: 
\begin{eqnarray}
	U_c(T) &=& 
	U^* e^{b(T-T_c)} + e^{bT} \int^T_{T_c}dT' A(T')e^{-bT'}
	\nonumber
	\\
	&=& \left[U^* + \frac{\gamma + \eta\sqrt{T_c}}{b}\right] e^{b(T-T_c)} 
- \frac{\gamma + \eta\sqrt{T}}{b} 
\nonumber
\\
&&+ \frac{\sqrt{\pi}(2\alpha b + \eta)}{2b\sqrt{b}} e^{bT}
[\mathrm{erf}(\sqrt{bT}) - \mathrm{erf}(\sqrt{bT_c})],
\nonumber
\\ \label{eq:PTL}
\end{eqnarray}
where $T{<}T_c$ 
and $\mathrm{erf}(x)$ is the error function defined by
\begin{equation}
	\mathrm{erf}(x) \equiv \frac{2}{\sqrt{\pi}}\int_0^{x} dt e^{-t^2}.
\end{equation}

In fact, $A(T)$ is sensitive to the set of data points that we use to obtain the 
best fit to Eq.~(\ref{eq:linearU}). 
Initially, we guess the PTL $U_c(T)$ and calculate $F(T,U)$ for the data points around the line, 
which, in turn, yields a new PTL $U_c(T)$ from the best fit to Eq.~(\ref{eq:linearU}).
We have repeated the procedure until the data points used for the fitting reasonably 
overlap with the resulting PTL.
All the fitting results presented in Fig.~\ref{fig:fA} are those obtained using the
self-consistent parameters.
As can be seen in Fig.~\ref{fig:fA}(b), the values of the self-consistent $A(T)$ are well
approximated by the expression derived from the Fermi-liquid theory.

\begin{figure}
	\includegraphics[width=0.40\textwidth]{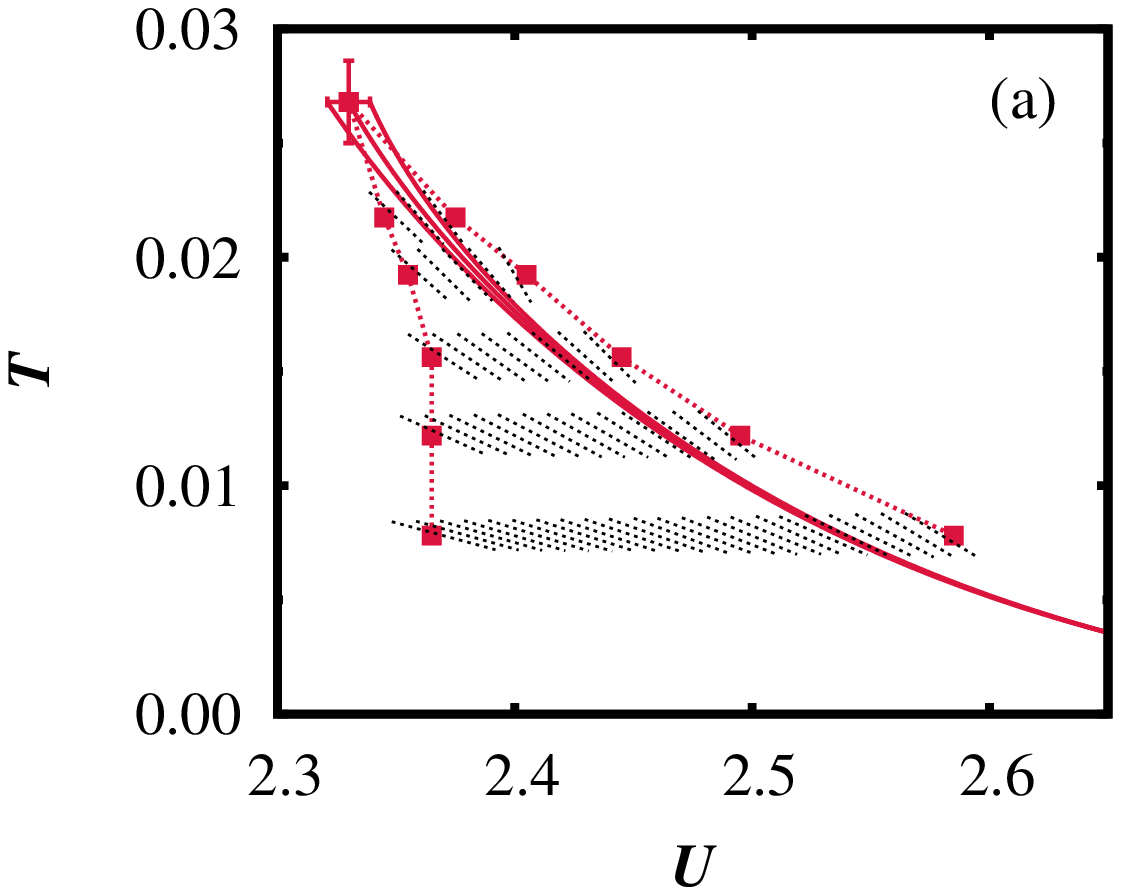}
	\includegraphics[width=0.40\textwidth]{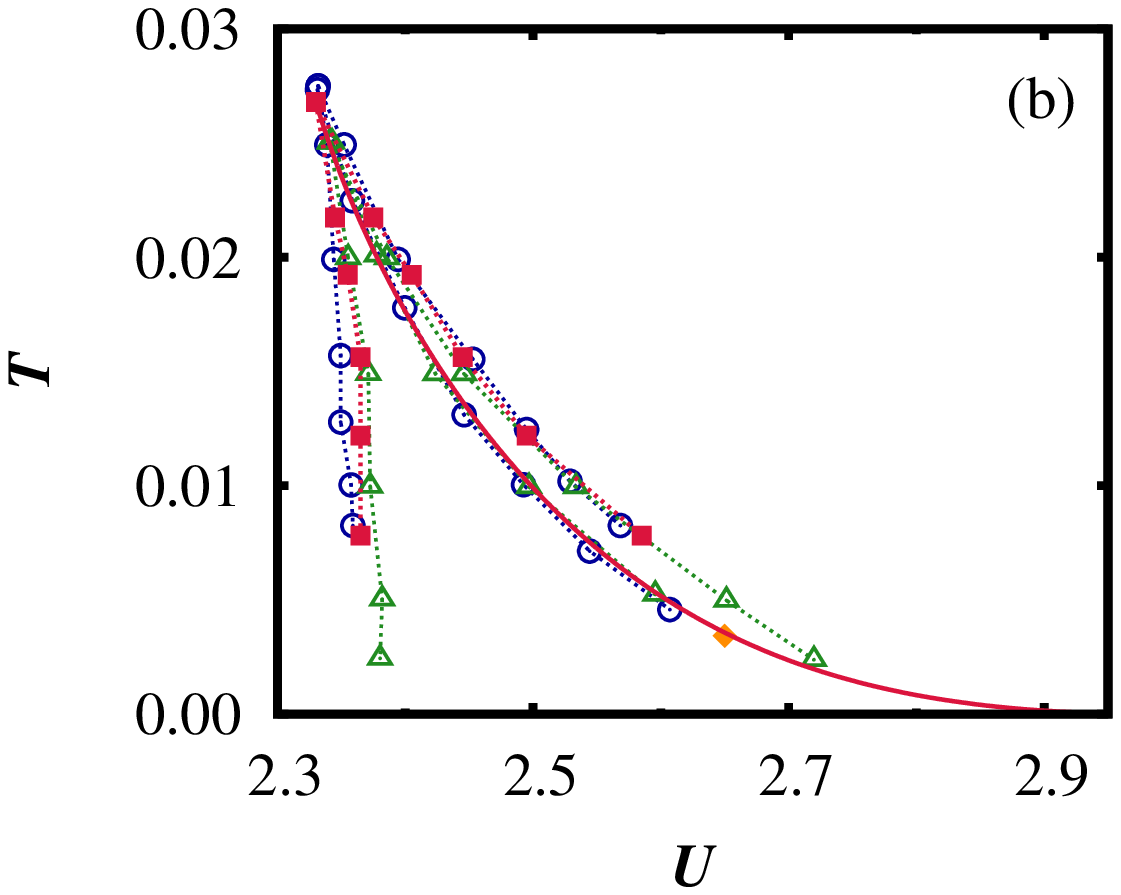}
	\caption{ (color online)
		(a) Phase transition line (PTL) of the infinite-dimensional
		Hubbard model  and (b) comparison with earlier DMFT results.
		In panel (a), short (black) dotted-line segments represent the lines with the slope
$1/F(T,U)$ at the point $(T,U)$. 
The solid lines are PTLs resulting from three different critical points that
are varied maximally within the numerical errors.
In panel (b), the transition interaction strengths $U_{c1}$ and $U_{c2}$, and the PTL obtained in this
work are denoted by (red) squares and a solid line, respectively.
The earlier results obtained by using the Hirsch-Fye quantum Monte Carlo~[\onlinecite{Blumer2002}], 
and exact diagonalization~[\onlinecite{Tong2001}] are
shown as (blue) empty circles and (green) triangles, respectively. 
The (orange) diamond represents an earlier CTQMC result ~[\onlinecite{Werner2006}].  
}
	\label{fig:phase}
\end{figure}

In Fig.~\ref{fig:phase}(a), we have shown the resulting PTL. 
Clearly, the slope of PTL
is in good agreement with the inverse of $F(T,U(T))$ near the PTL, 
which demonstrates the full self-consistency of the PTL.
We have also examined the dependence of the PTL on the variation in the position
of the critical point.
We have obtained PTLs for different critical points that were varied maximally
within the numerical errors.
The resulting PTLs shown in Fig.~\ref{fig:phase}(a) show little difference
particularly at low temperatures, which indicates that the PTLs estimated at low
temperatures are robust against any small variation in the position of the critical point. 

In Fig.~\ref{fig:phase}(b), we have also compared the PTL obtained in this work with those from earlier DMFT works.
The comparison shows that our results are fairly consistent with earlier results
up to the lowest temperature that the earlier works examined.
Particularly, the agreement with the results obtained from Hirsch-Fye
QMC method~\cite{Blumer2002} implies that our method is accurate enough to reproduce the 
low-temperature transition nature without any knowledge of the preceding zero-temperature
results.
Notably, our estimated PTL shows excellent agreement with the
result obtained from earlier CTQMC calculation at the temperature $T {\approx} 0.0034$~\cite{Werner2006}, which is much lower than the lowest
temperature $T{=}1/128$ for which we were able to obtain raw data from the CTQMC method, indicating the
efficiency of our method.

\subsection{Transition Interaction Strength at Zero Temperature}
\begin{figure}
	\includegraphics[width=0.40\textwidth]{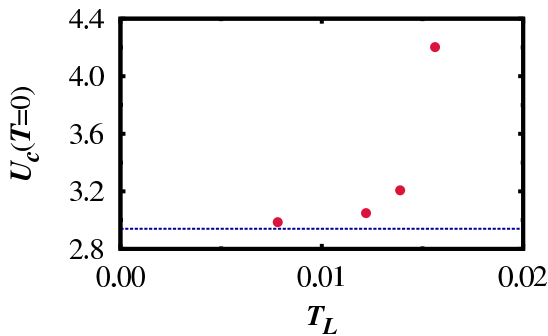}
	\caption{ (color online)
		$U_c(T{=}0)$ obtained from the CTQMC data in the temperature range of $T_L{<}T{<}T_H$.  
		We have fixed $T_H{=}1/46$.
		The (blue) dotted line indicates the zero-temperature NRG result
		$U_c(T{=}0){=}2.94$~[\onlinecite{Bulla1999}].}
	\label{fig:UcT0}
\end{figure}

From the PTL in Eq.~(\ref{eq:PTL}), we can easily estimate the 
transition interaction strength at zero temperature, which yields $U_c(T{=}0) {\approx} 3.04$;
this value is slightly higher than the zero-temperature NRG result $U_c {\approx} 2.94$~\cite{Bulla1999}.
We have also examined the dependence of the estimated $U_c(T{=}0)$
on the temperature range of the CTQMC data used in the procedure.
In determining the PTL, we have used the CTQMC data in the range of $T_L{<}T{<}T_H$,
where we have fixed the upper-limit temperature as $T_H{=}1/46$ and varied the lower-limit temperature $T_L$.
Figure~\ref{fig:UcT0} shows $U_c(T{=}0)$ as a function of $T_L$.
As $T_L$ is lowered, the estimated $U_c(T{=}0)$ shows a monotonic decrease and rapidly approaches the
zero-temperature NRG result, which is denoted by the dotted line in the figure. 
Such a rapid monotonic approach implies that the CTQMC results at finite temperatures are fully
consistent with those from the zero-temperature approaches.
Here, we should note that our method does not use any 
knowledge of previous zero-temperature results; this implies that our
results in the zero-temperature limit provide an independent check on the
zero-temperature result.

\subsection{Next-order Correction}

Finally, we check out the validity of the assumption in Eq.~(\ref{eq:linearU}) that
$F(T,U)$ is linear in $U$.
To include the next-order correction, we try the nonlinear function
\begin{equation}
	F(T,U) = \frac{A(T) + bU(1+\zeta U)}{1 + 2\zeta U},
	\label{eq:nonlinear}
\end{equation}
which includes a second-order correction in $U$ and allows an analytic
solution for the differential equation in Eq.~(\ref{eq:diffeq}).
Although this form can include the curvature of data, the singularity in
the denominator, which is introduced to allow an analytic solution for the differential
equation, limits the range of the parameter $\zeta$.
We have varied $\zeta$ in the range between -0.1 and 0.1 and found that $U_c(T{=}0)$
changes only by an amount 0.01, indicating that our low-temperature results are also
robust against the small corrections to Eq.~(\ref{eq:linearU}) arising from next-order terms in $U$.

\section{Summary}
In summary,
we have examined the Mott transition of the Hubbard model at finite temperatures 
in infinite dimensions by using the DMFT with CTQMC method being an impurity solver.
We have measured the double occupancy and the energy density, which yields a critical point, as
well as a coexistence region, in the phase diagram. 
We have determined the PTL of the first-order Mott transition by integrating the differential equation of the transition interaction strength.
The PTL constructed in this way has been shown to be in good agreement with
earlier results. 
We have also shown that higher-order corrections do not have much effect on the
low-temperature PTL.

\acknowledgments
We thank Prof.\ P.\ Werner and Dr.\ H.\ Lee for helpful discussions on the CTQMC
method.
This work was supported by National Research Foundation of Korea 
through grants 2008-0061893, 2013R1A1A2007959 (A.J.K and G.S.J.), and
2012R1A2A4A01004419 (M.Y.C.).

\end{document}